\documentclass[
    amsmath,
    amssymb,
    aps,
    floatfix,
    twocolumn
]{revtex4-2}

\usepackage[utf8]{inputenc}
\usepackage[colorlinks=true,citecolor=blue,allcolors=blue]{hyperref}
\usepackage{amsmath,amsthm,amsfonts,amssymb,amscd}
\usepackage{graphicx}
\usepackage{lastpage}
\usepackage{enumerate}
\usepackage{mathrsfs}
\usepackage{bm}
\usepackage{braket}
\usepackage[version=4]{mhchem}
\usepackage{blindtext}
\usepackage{natbib}
\usepackage{url}

\newcommand\mati{\begin{matrix}}
\newcommand\matf{\end{matrix}}
\newcommand\bmati{\begin{bmatrix}}
\newcommand\bmatf{\end{bmatrix}}
\newcommand\pmati{\begin{pmatrix}}
\newcommand\pmatf{\end{pmatrix}}

\newcommand{\pushright}[1]{\ifmeasuring@#1\else\omit\hfill$\displaystyle#1$\fi\ignorespaces}
\newcommand{\pushleft}[1]{\ifmeasuring@#1\else\omit$\displaystyle#1$\hfill\fi\ignorespaces}

\newcommand{\quotes}[1]{``#1''}

\begin{document}

\author{
    Joshua Forer$^{1}$,
    Jeoffrey Boffelli$^{2}$,
    Mehdi Ayouz$^3$,
    Dávid Hvizdoš$^4$,
    Viatcheslav Kokoouline$^5$,
    Ioan F. Schneider$^{2}$,
    Chris H. Greene$^{4}$
}

\affiliation{
  $^1$Columbia Astrophysics Laboratory, Columbia University, New York, New York 10027, USA \\
  $^2$Laboratoire Ondes et Milieux Complexes, CNRS, Universite\'e le Havre Normandie, 53 rue Prony, Le Havre, 76058, France.
\\
  $^3$Universit\'e Paris-Saclay, CentraleSup\'elec, Laboratoire de G\'enie des Proc\'ed\'es et Mat\'eriaux, 91190, Gif-sur-Yvette, France. \\
  $^4$Department of Physics and Astronomy and Purdue Quantum Science and Engineering Institute, Purdue University, West Lafayette, Indiana 47907, USA \\
  $^5$Department of Physics, University of Central Florida, 32816, Florida, USA \\
  \looseness=-6 
}

\title{Dissociative recombination of \ce{CF+}}

\date{\today}

\begin{abstract}
  This work presents our theoretical study of the dissociative recombination (DR) of the closed-shell diatomic system \ce{CF+} based on an approach recently applied to the \ce{CH+} molecule.
Our extended treatment uses the UK R-matrix theory and the multichannel quantum-defect theory procedure to uniformly resolve the direct and indirect mechanisms of DR while bypassing explicit dissociative state and electronic coupling calculations.
The theoretical results exhibit good overall agreement with previous experimental measurements.
At lower scattering energies, good agreement is found only if the rotational structure of the ion is included and the theoretical cross sections are averaged over initial rotational levels corresponding to the temperature at which the experimental measurements were made.
At higher scattering energies, our rotationally resolved results are very similar to those obtained without including the ion's rotational structure.
\end{abstract}

\maketitle

{\it Relevance to fluorine-containing molecular plasmas.} Fluorine-containing molecular species play a distinct role, compared to other species, in the chemistry of gases that are subjected to ultra-violet (UV) radiation, such as in the interstellar medium (ISM)\cite{neufeld2009chemistry,dagdigian2019calculation}, owing to a combination of the high ionization potential of atomic fluorine and the large dissociation energies of HF and \ce{HF+}.
The ubiquitous hydrogen atoms shield atomic fluorine in interstellar clouds from ionizing UV radiation because the ionization potential of F is higher than that of H, which results in a mostly neutral population of atomic fluorine. Therefore, the fluorine chemical network in the ISM is quite simple\cite{neufeld2009chemistry}, when compared to other chemical networks involving significant populations of ionized atoms, and can be used to benchmark astrophysical models of interstellar clouds. Atomic fluorine can collide with \ce{H2} to form \ce{HF}, which then reacts with \ce{C+} to produce \ce{CF+}. The \ce{CF+} ion, in turn, can be destroyed by dissociative recombination (DR) with free electrons, producing atomic fluorine and closing the network loop:
\begin{equation}
\begin{gathered}
\label{eq:F_network}
\ce{F + H2 -> FH + H},\\
\ce{HF + C+ -> CF+ + H},\\
\ce{CF+ + e- -> C + F}.
\end{gathered}
\end{equation}
The above reactions are the major steps in the fluorine cycle in the ISM, which makes \ce{HF} and \ce{CF+} the first and second most abundant fluorine-containing molecules in that environment, where their relative abundances are determined by the rate coefficients of the two last reactions in (\ref{eq:F_network}). To understand fluorine chemistry in the ISM and the above-mentioned benchmarking of the astrophysical models by comparing them with observations, it is essential to have accurate rate coefficients for the above reactions.

One astrophysical model~\cite{neufeld2009chemistry} of the fluorine network predicted \ce{CF+} abundances that are more than an order of magnitude larger than the abundances inferred from observations \cite{neufeld2006discovery,neufeld2009chemistry}. It is likely that one or both rate coefficients for the two last reactions in (\ref{eq:F_network}) used in the model are not accurate.  The rate coefficient for the second reaction in (\ref{eq:F_network}) was not experimentally inferred. Different theoretical values, obtained by a modified Langevin method~\cite{neufeld2009chemistry} and a quasi-classical trajectory method~\cite{denis2018study} differ from each other by about an order of magnitude.  The rate coefficient for the third reaction (DR) in (\ref{eq:F_network}) used in the model, $5.2\times 10^{-8}(T/300~\mathrm{K})^{-0.8}$~cm$^3$/s, was derived from the ASTRID and CRYRING storage-ring experiments of \citet{novotny2005dissociative}. In those experiments, the rovibrational temperature of the \ce{CF+} ions were very high, around 6000~K, according to the study~\cite{novotny2005dissociative}. Therefore, the DR rate coefficient derived from the experiments may not be appropriate to model the cold (30--300~K) interstellar environment where the ions are all expected to be in one of the lowest rovibronic (rotational, vibrational, and electronic) state. The present study is devoted to an accurate determination of the DR rate coefficient for \ce{CF+}.

\ce{CF+} was first detected by \citet{neufeld2006discovery} and later found in different ISM environments in the Milky Way, as well as in other galaxies \cite{kalenskii2010spectral,kalenskii2010spectralb,gong2024first} (see further references in the cited works). With its relatively simple chemical network in the ISM, the detection of \ce{CF+} can help constrain astrochemical models, particularly those in which \ce{C+} plays a significant role.
In addition to the experimental data, there are theoretical DR cross sections reported in the same study by \citet{novotny2009dissociative}. The theoretical results agree very well overall with the experiment. We note that no thermally averaged rate coefficients were computed from the DR cross sections obtained by the authors of that study. However, because the theoretical (taking into account the anisotropic collisional energy distribution of the experiment) rate coefficient agrees well with the experimental coefficient in the interval of energies 0.001-0.6~eV, the thermal rate coefficients should also agree with each other for temperatures of 10--3000~K.

For accurate modeling of the cold interstellar environment, one would need rate coefficients obtained for the lowest rotational states of \ce{CF+}. As mentioned above, the storage ring experiments were performed at a high rovibrational temperature of the ions. The theoretical calculations did not account for the rotational structure and, therefore, cannot provide the information about the dependence of the rate coefficients on rotational excitation of \ce{CF+}. In addition, the previous theory did not account for the effect of the electronic Rydberg resonances associated to the first excited state, a$^3\Pi$, of the ion. If the incident electron is captured in one of these resonances, it should enhance the DR cross section. Therefore, the inclusion of the coupling of the incident electron with a$^3\Pi$ Rydberg states may be important at electron energies at and above 1~eV, where these resonances should appear in the DR spectrum.

We have recently~\cite{jiang2019cross,jiang2021theory,forer2023unified} developed a theoretical method that allows us to account for the direct and indirect (via rotational, vibrational, or electronic resonances) mechanisms~\cite{bardsley68} of DR, while accounting for the rovibronic structure of the target ion and the coupling of the incident electron with a large number of rovibronic ionic states to the dissociative continuum of the neutral system. The present study applies the method to the DR of \ce{CF+}. Accurate DR cross sections are also important for understanding and modeling processes in the plasma etching of semiconductors. Fluorocarbon species are commonly used for etching; after a radio-frequency discharge, a molecular plasma is formed in which \ce{CF+} is one of the major ions \cite{goyette2000ion}, with DR being the major mechanism of plasma neutralization.

Because the theoretical method employed here has already been described by \citet{forer2023unified},
the following discussion presents only the details specific to the \ce{CF+} DR process.

\begin{figure}
    \includegraphics[width=\linewidth]{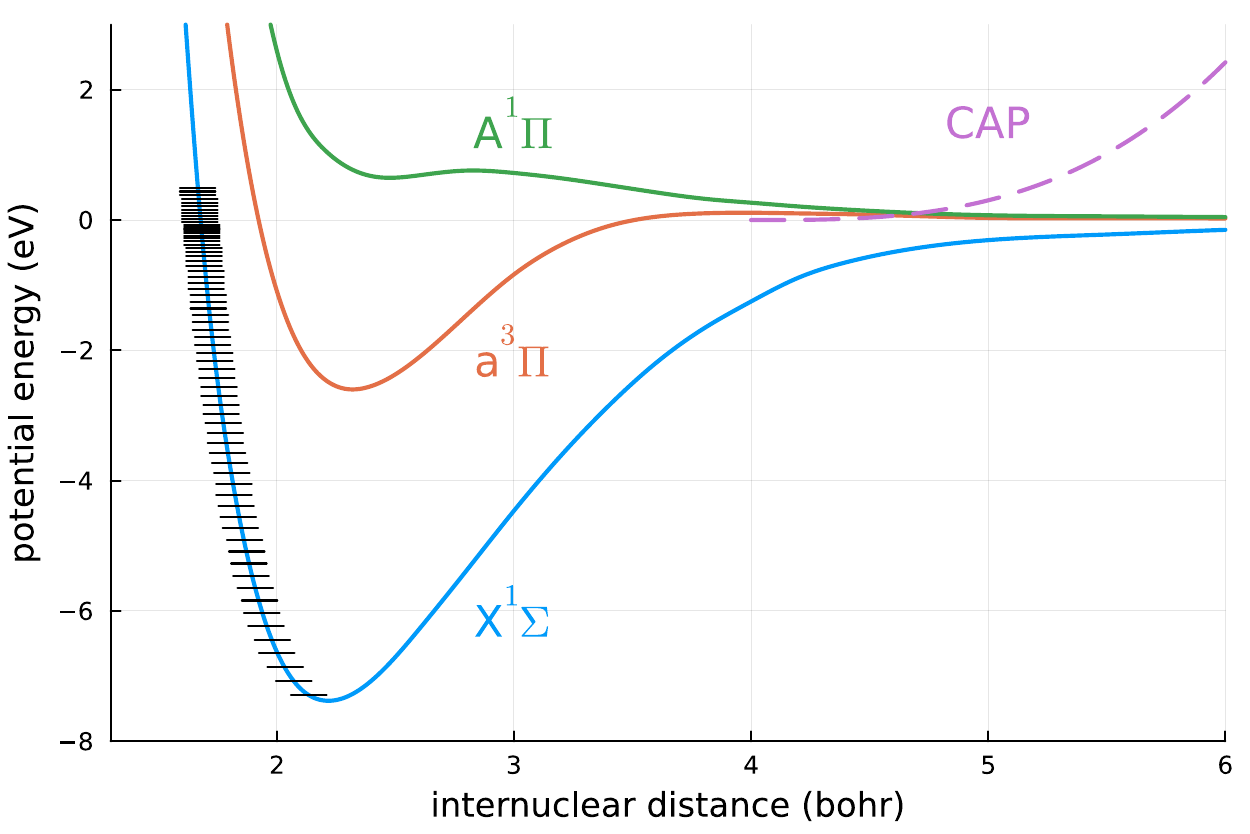}
    \caption{
      The potential energy curves for the lowest three electronic states of \ce{CF+}, obtained using the work of \citet{petsalakis2000electronic}.
      The dashed line represents the complex absorbing potential used for the ionic ground state in the vibrational frame transformation.
      The straight lines through the repulsive part of the X$^1\Sigma$ curve represent the real part of the energy of the vibrational levels used in our approach. The dissociation limit for the displayed potential energy curves is 0~eV.
    }
    \label{fig:PEC}
\end{figure}

{\it Electronic states of} \ce{CF+}{\it and} \ce{CF}.
Born-Oppenheimer energies of several low-lying singlet and triplet states of the \ce{CF+} ion were determined as functions of the internuclear distance $R$ by \citet{petsalakis2000electronic} using the multireference configuration interaction method. The energies of the three lowest states (X$^1\Sigma^+$, a$^3\Pi$, A$^1\Pi$) used in this study are shown in Fig.~\ref{fig:PEC}. The a$^3\Pi$ state has a relatively shallow minimum, when compared to the ground state. The A$^1\Pi$ state has an even shallower minimum, which lies above the \ce{C+(^2P) + F(^2P)} dissociation limit, thereby possessing only predissociated vibrational levels.
The \ce{CF} potential energy curves were later computed by the same authors \cite{petsalakis2011theoretical}. While the present study does not require an explicit form of the potential curves of the neutral molecule, we note here that the lowest \ce{CF} curves of the $^2\Sigma^+$, $^2\Sigma^-$, $^2\Pi$ and $^2\Delta$ symmetries cross the ground state of the ion near its equilibrium geometry. In particular, the potential curve of the lowest state of $^2\Sigma^-$ symmetry crosses the ionic ground state almost exactly at the equilibrium and one of the $^2\Sigma^+$ curves of the neutral crosses the ionic ground state slightly to the left from the equilibrium. This means that the direct DR mechanism, whereby \ce{CF} rovibrational Rydberg resonances are not excited, is expected to have a significant contribution.

{\it Ab initio \ce{CF+ + e-} scattering calculations}.
The first step in the approach involves quantum chemistry calculations of the target (\ce{CF+}).
This is followed by building the scattering matrix ($S(R)$) in the Born-Oppenheimer approximation with a fixed-nuclei R-matrix method implemented in the {\tt UKRMol+} suite~\cite{mavsin2020ukrmol+}, called with the UKRmol scripts~\cite{houfek2024ukrmol}, for several values of the internuclear coordinate $R$.
Before performing the electron-scattering calculations, the electronic states of the target are constructed with the Psi4 quantum chemistry package~\cite{turney2012psi4} in the $C_{2v}$ point group with the {\tt cc-pVTZ} basis set, using 2 frozen orbitals (1--2$A_1$) and 8 active orbitals (3--6$A_1$, 1--2$B_2$, 1--2$B_2$).
Scattering calculations are performed with a close-coupling expansion, using the target's active space supplemented with 6 virtual orbitals (7--9$A_1$, 3$B_1$, 3$B_2$, 1$A_2$), as described in the work of \citet{mavsin2020ukrmol+}.
The two lowest electronic states of the ion are separated by about 5~eV at the equilibrium distance (2.216~bohr).
This excited state is low enough in energy to generate a Rydberg series of electronic resonances that affect low-energy electron scattering.
These Rydberg states provide an indirect pathway to the dissociative states of the neutral molecule (\ce{CF}).
Although we mentioned in the introduction that we expect direct DR to contribute significantly, both mechanisms can occur simultaneously and can interfere.

The body frame reaction ($K(R)$)-matrices are obtained in the same manner as in the previous study~\cite{forer2023unified}: a K-matrix is obtained for each geometry with the R-matrix calculations, using an R-matrix radius of 13~bohr, and is evaluated at $\sim5\times10^{-3}$~eV above the ground state with channels attached to the excited states treated as {\it strongly closed}.
The scattering calculation at the internuclear coordinate $R$ at equilibrium $R=R_e$  was performed with the origin for the partial-wave expansion at  the target's center of charge (at a distance $d_{C_e}$ from the center of charge to the C nucleus).
Calculations for all other values of $R$ were performed with the origin situated at a distance $d_C$ from C proportional to $d_{C_e}$ and $R$: $d_C=d_{C_e} R/R_{e}$.
The center-of-charge frame is preferable because scattering calculations for dipolar targets converge faster with respect to the partial wave sum of the scattering electron.
We identify the electronic state of the ion with $n$, the orbital angular momentum quantum number of the electron with $l$, and the projection of $\vec{l}$ on the molecular axis with $\lambda$.
Performing the closed-channel elimination on the R-matrix, as described by \citet{hvizdovs2023bound}, produces resonances in the K-matrices, so care must be taken to sample K-matrices at energies where the energy dependence is smooth.
We consider 9 partial waves ranging from $l=0$--$2$ in the scattering calculations, each attached to the $X^1\Sigma^+$, a$^3\Pi$, and A$^1\Pi$ states of \ce{CF+}.
The K-matrices are then converted to S-matrices via the same method presented in the previous study~\cite{forer2023unified}.

{\it Scattering matrix with rovibronic channels.}
The next step in our approach, the rovibrational frame transformation, is carried out in two steps: an energy-independent vibrational frame transformation followed by a rotational frame transformation.
Compared to the previous study~\cite{forer2023unified}, the frame transformation has two differences: (i) we solve the vibrational Schr\"odinger equation using a basis of B-splines~\cite{Forer_BSPRVSE_2024} instead of a discrete variable representation (DVR) method and (ii) the complex absorbing potential's (CAP) imaginary component takes the cubic form
\begin{equation}
  V_{CAP} =
  \begin{cases}
    -2i\eta (R-R_0)^3 & R > R_0
    \\
    0 & R \le R_0
  \end{cases}
  ,
  \label{eqn:CAP}
\end{equation}
instead of the previously used exponential form.
In (\ref{eqn:CAP}), $R_0$ is the starting point of the CAP (4.0 bohr),  $\eta$ is the CAP strength (0.005561 atomic units), and $L$ is the CAP length (2.0 bohr).
The vibrational quantum number $v$ runs from $0$--$69$.
Similarly to previous studies~\cite{kokoouline2005theoretical,vcurik2008rates,forer2023unified}, bound states and quasi-continuum states must be represented by vibrational wave functions used in the transformation.
Quasi-continuum states are chosen to satisfy outgoing-wave boundary conditions, which represent outgoing dissociative flux.
The resulting vibronic S-matrices,
now with no $R$-dependence, are expressed in a basis of 630 vibronic channels.

The vibronic S-matrices undergo a rotational frame transformation as described in the previous study \cite{forer2023unified}.
Channels of the resulting S-matrices are numerated by  {\it rovibronic} quantum numbers, $n$-the index of the electronic state of the ion, $v,j$- the vibrational and rotational quantum states in the electronic state $n$.
If one is not interested in the rotational structure of the final cross sections (for example, to compare with experimental data where the rotational structure was not resolved), the rotational frame transformation can be omitted.

 The energy of the initial state of \ce{CF+}, $E_{nv(j\mu)}$, and the energy of the incident electron with respect to $E_{nv(j\mu)}$, $E_\text{el}$, define the total energy of the system, $E_\text{tot}$:
\begin{equation}
  E_\text{tot} = E_{nv(j\mu)} + E_\text{el}.
  \label{eqn:Eto}
\end{equation}
We use parentheses around the quantum numbers $j$ and $\mu$ to indicate the optional inclusion of the rotational resolution from the rotational frame transformation ; the quantum numbers $j$ and $\mu$ are absent if it is bypassed.
For total energies where some scattering channels are closed, the closed-channel elimination procedure, (CCEP) equation (2.50) in the work of \citet{aymar1996multichannel},
is applied \cite{kokoouline2003unified,forer2023unified}.

The CCEP produces the scattering matrix -- we call it the physical S-matrix -- $S^{phys}(E_\mathrm{tot})$, which is strongly energy-dependent due to the Rydberg resonances associated with closed rovibronic (or vibronic, if the rotational structure is neglected) channels. The open and closed channels of the S-matrix with the rotational structure are labeled with quantum numbers $\{n,v,j,\mu,l\}$, mentioned above. In addition, there is a quantum number of the total angular momentum $J$ and its projection of the molecular axis, over which the S-matrix is diagonal. We use the notation $S^{J,phys}_{nvj\mu l,n'v'j'\mu'l'}$ for the S-matrix elements.  The channels of the S-matrix without the rotational structure are enumerated by the set of quantum numbers $\{n,v,l,\lambda,\Lambda\}$, where $\Lambda$ is the projection of the angular electronic angular momentum (target + incident electron) on the molecular axis.   We use notation $S^{\Lambda,phys}_{n'v'l'\lambda',nvl\lambda}$ for these elements.
Here and throughout, primed quantities refer to quantum numbers or indices of final states, while their unprimed counterparts refer to initial states.

{\it DR cross sections and rate coefficients.}
Equipped with the physical (ro)vibronic S-matrices, we can compute the total DR cross section from an initial state of the system, $nv(j\mu)$:
\begin{widetext}
\begin{gather}
    \sigma_{nv}(E_\text{el}) = \frac{\pi}{2m_eE_\text{el}} \sum\limits_{l\lambda} \big[ 1 -  \sum\limits_{l'\lambda'} \sum\limits_{n'v'} S^{\Lambda,phys}_{n'v'l'\lambda',nvl\lambda}S^{\Lambda,phys\ddagger}_{nvl\lambda,n'v'l'\lambda'} \Big],
    \label{eqn:xs_norot}
    \\
    \sigma_{nvj\mu}(E_\text{el}) = \frac{\pi}{2m_eE_\text{el}} \sum\limits_J \frac{2J+1}{2j+1} \sum\limits_{l} \Big[ 1 - \sum\limits_{l'} \sum\limits_{n'v'j'\mu'} S^{J,phys}_{n'v'j'\mu'l',nvj\mu l}S^{J,phys\ddagger}_{nvj\mu l,n'v'j'\mu'l'} \Big].
    \label{eqn:xs_rot}
\end{gather}
\end{widetext}
where $m_e$ is the mass of the electron.
Calculations with rotational resolution use (\ref{eqn:xs_rot}), while those without it use (\ref{eqn:xs_norot}).

\begin{figure}
    \includegraphics[width=\linewidth]{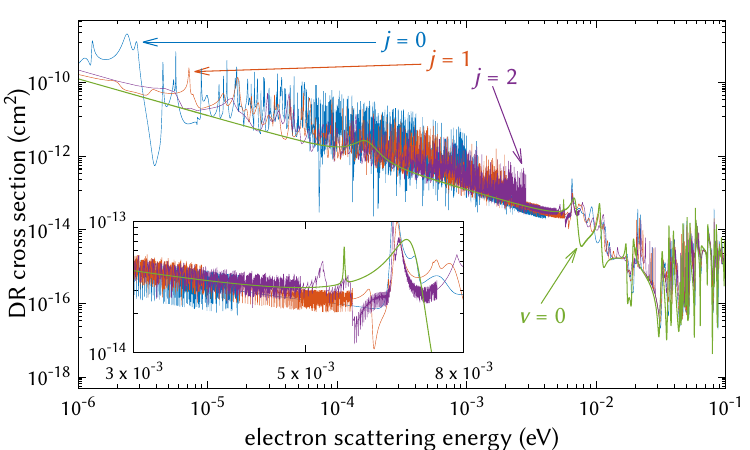}
    \caption{
      \ce{CF+} DR cross sections.
      Cross sections with rotational structure are shown from the three lowest states of the ion, $j=0,1,2$.
      Cross sections from the ground vibronic state, without the rotational structure, are also shown $v=0$.
    }
    \label{fig:DR_XS}
\end{figure}

{\it Results.}
Fig. \ref{fig:DR_XS} presents the calculated DR cross sections for different rovibronic levels ($n=1$; $v=0$; $j=0$, 1, 2) of \ce{CF+} and one calculation without rotational resolution ($n=1$; $v=0$).
We restrict the calculation here to low-scattering energy ($10^{-4}-10^{-1}$~eV) to stress the rotational effects on the cross section.
The overall behavior of the cross sections at the lowest energies is $1/E_\text{el}$.
Many resonances are present and can be grouped into two different classes depending on their origin.
The first class consists of sharp and narrow predissociated Rydberg resonances driving the indirect mechanism.
Some of these are shown in the inset of Fig. \ref{fig:DR_XS}.
The second class are broad valence resonances driving the direct mechanism.

\begin{figure}
    \centering
    \includegraphics[width=0.48\textwidth]{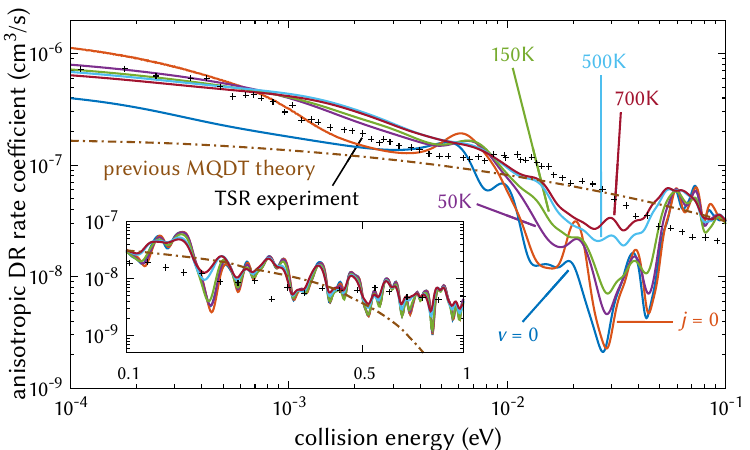}
    \caption{
      \ce{CF+} rate coefficients.
      Solid curves indicate results from the present theory, the dashed curve represents the results of the an MQDT calculation, and the points are experimental results obtained at the TSR.
      The MQDT calculation and TSR results are from the same study~\cite{novotny2009dissociative}.
      Two of the solid curves are state-selected data convolved with experimental parameters, but not thermally averaged over initial states: $v=0$ and $j=0$.
      The other solid curves represent data convolved with experimental parameters and averaged over populations of initial states at the indicated rotational temperatures.
      The inset shows the same data in the energy range 0.1--1~eV.
    }
    \label{fig:RateAniso}
\end{figure}
Fig. \ref{fig:RateAniso} compares the results obtained, with and without rotational resolution ($v=0$ and $j=0$, respectively), a previous MQDT calculation, and experimental measurements made at the TSR --- the latter two are from the work of \citet{novotny2009dissociative}.
Various convolutions of our rotational results are also displayed in Fig.~\ref{fig:RateAniso}.
The cross sections are convolved with an anisotropic Maxwell-Boltzmann distribution, which accounts for different populations of initial states of \ce{CF+} as well as the longitudinal and transverse energy spreads in the TSR experiment, as described by equations 5 and 6 in the work of \citet{santos07} and \citet{forer2024kinetic}.
Our $v=0$ and $j=0$ results are only convolved with the TSR collisional energy spreads; these results remain state-selected.
The remaining curves were averaged over the initial states of \ce{CF+} at several rotational temperatures, up to 700~K.
Measurements were made at the TSR experiment~\cite{novotny2009dissociative} while the ions cooled from $\sim$ 6000~K to 700~K, which suggests that a convolution of our results at only 700~K may not be entirely appropriate for comparison to the TSR results.

As is evident in Fig.~\ref{fig:RateAniso}, the present thermally averaged results approach the experimental data in the range of collision energies between $10^{-2}$ and $6\times10^{-2}$~eV as the rotational temperature of the thermal average increases.
The discrepancy between our $v=0$ results and the TSR results seems to diminish as more rotational states are included at rotational temperatures approaching those of the experiment.
Our non-thermally averaged rovibronic ($j=0$) and vibronic ($v=0$) results are most different at lower collision energies, but are much more similar above $\sim6\times10^{-2}$~eV.
Fig. \ref{fig:RateAniso} gives an idea of the ranges where rotation can be reasonably neglected to study DR in the case of \ce{CF+}.
As can be seen in the inset, the results are overall very similar above $10^{-1}$~eV.
As the collision energy approaches 1~eV, the theoretical results of \citet{novotny2009dissociative} drop off significantly, while our results continue to find good agreement with their experimental measurements.
This may be due to the lack of the a$^3\Pi$ state of \ce{CF+} in their theoretical mode, as mentioned in the introduction.

\begin{figure}
    \centering
    \includegraphics[width=0.48\textwidth]{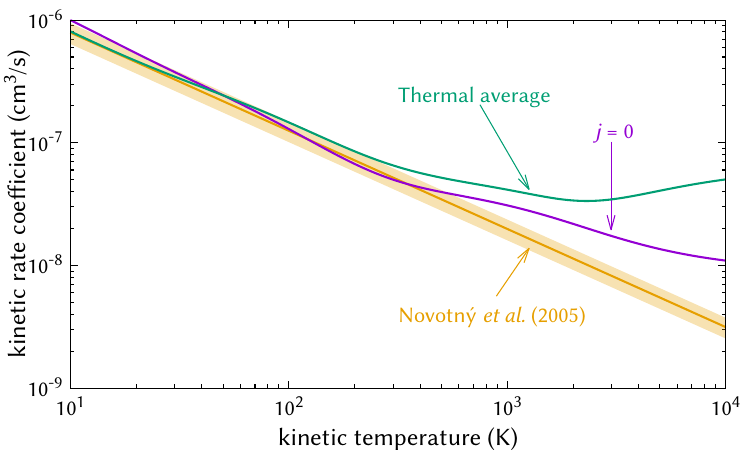}
    \caption{
      DR thermally averaged rate coefficients. The \quotes{$j=0$} line is the rate coefficient obtained assuming that the \ce{CF+} ions are in their ground rovibrational state and that the electron energy distribution follows Maxwell-Boltzmann statistics; the \quotes{Thermal average} line is obtained assuming a Maxwell-Boltzmann distribution over initial rovibrational states of the ions at the same temperature as the free electrons, i.e. averaged over the thermal ensemble of the ions. The experimentally determined rate coefficient, with error bars, of \citet{novotny2005dissociative} is also shown as a function of temperature.
    }
    \label{fig:thermalRates}
\end{figure}

Figure \ref{fig:thermalRates} compares the state-selected kinetic rate coefficients obtained in this study to those reported by \citet{novotny2005dissociative}  and used in the astrophysical model  of the fluorine chemistry in the ISM \cite{neufeld2009chemistry}. Our results are very similar in the temperature range 0-40~K.

{\it In summary,} we would like to stress the following key conclusions of the study.
\begin{itemize}
\item
The present theoretical results are in excellent agreement with the TSR and ASTRID storage ring data~\cite{novotny2009dissociative} of \ce{CF+} DR. Taking into account the anisotropic electron energy distribution in the experiments, the present theoretical rate coefficient (Fig.~\ref{fig:RateAniso}) agrees very well with the experiment at lower temperatures.
\item
The good agreement with two experiments demonstrates the validity and performance of the theoretical approach to model the DR process.
\item
We have performed complete DR calculations with and without accounting for the rotational structure of the target ion and analyzed the obtained results with the idea of understanding when the inclusion of the rotational structure is needed and when it can be neglected.
\item
The study has illustrated the importance of rotational resonances in DR spectra, especially at low energies. This could be a quite general conclusion for many target ions for which the coupling of the incident electron with rotational motion of the target ion is significant: such a coupling produces rotational Rydberg resonances if the corresponding rotational channel is closed and, therefore, increases the time during which the ion-electron system can undergo the dissociation process.
The term \quotes{low energy} should be understood as the energy below the excitation threshold of the lowest rotational states of the target ion.
The rotational effects are effectively averaged out at higher collision energies.
\item
The rotational structure of the ion is important to include for low-energy DR, as evident in  Fig.~\ref{fig:RateAniso} for the low-energy $v=0$ and $j=0$ rate coefficients.
The figure also illustrates how one may well ignore the rotational resolution at higher scattering energies, given the similarity in the $v=0$ and $j=0$ rate coefficients above $6\times10^{-2}$~eV.
 \item
The study has also demonstrated the role of rotational and, in general, rovibronic structure in interpretation of experimental results from storage rings,  like those from the TSR and ASTRID~\cite{novotny2009dissociative}.
DR cross sections (and anisotropic rate coefficients) as functions of collision energy tend to show typical drops when rotational, vibrational, or electronic thresholds are opening: If the coupling between the incident electron and the corresponding motion of the ion (rotational, vibrational, or electronic) is strong, below such threshold energies electron-ion collisions lead to a large probability of DR, while at energies above the thresholds, the collisions lead to the excitation -- rotational, vibrational, or electronic -- of the ion.
\end{itemize}

\section{Acknowledgements}

We acknowledge support from NSF award nos. 2303895, 2102187, and 2102188.
M. A. and V. K. also  acknowledge support from the program “Accueil des chercheurs étrangers” of CentraleSupélec.
J. B. and I. F. S. acknowledge support from the French consortiums LabEx EMC3, FR-IEPE, and PN-PCMI.


\providecommand{\noopsort}[1]{}\providecommand{\singleletter}[1]{#1}%

\end{document}